\documentclass[twocolumn,superscriptaddress,nobibnotes,nofootinbib,floatfix,aps,prb,citeautoscript,reprint]{revtex4-1}

\usepackage{graphicx}
\usepackage{epsfig}
\usepackage{subfigure}
\usepackage{color}
\usepackage{latexsym}
\usepackage{amsmath,bm}
\usepackage{natbib}

 \newcommand{\lpmcn}{Universit\'e
  de Lyon, F-69000 Lyon, France and LPMCN, CNRS, UMR 5586,
  Universit\'e Lyon 1, F-69622 Villeurbanne, France}
\newcommand{\lsi}{Laboratoire des Solides Irradi\'es and ETSF, \'Ecole
  Polytechnique, CNRS, CEA-DSM, 91128 Palaiseau, France}
\newcommand{\ub}{Department of Physics, Universit\"{a}t Basel,
  Klingelbergstr. 82, 4056 Basel, Switzerland}

\begin{document}

\title{Low-energy silicon allotropes with strong absorption in the visible for photovoltaic applications}

\author{Silvana Botti}
\affiliation{\lsi}
\affiliation{\lpmcn}
\author{Jos\'e A. Flores-Livas}
\affiliation{\lpmcn}
\author{Maximilian Amsler}
\affiliation{\ub}
\author{Stefan Goedecker}
\affiliation{\ub}
\author{Miguel A.L. Marques}
\affiliation{\lpmcn}

\date{\today}

\begin{abstract}
We present state-of-the-art first-principle calculations of the
electronic and optical properties of silicon allotropes with
interesting characteristics for applications in thin-film solar
cells. These new phases consist of distorted sp$^3$ silicon networks
and have a lower formation energy than other experimentally produced
silicon phases. Some of these structures turned out to have
quasi-direct and dipole-allowed band gaps in the range 0.8--1.5\,eV,
and to display absorption coefficients comparable with those of
chalcopyrites used in thin-film record solar cells.
\end{abstract}
\maketitle

If Nature chose carbon as the scaffolding of life, mankind chose
silicon as the building block of much of the high-technology of today.
In fact, its advantages are many: silicon is the second most abundant
element in the Earth's crust, its processing is well controlled at the
industrial level, and its electronic properties and defect physics are
theoretically well understood.  As an elemental substance, it is an
intrinsic semiconductor that can readily be $p$-- and $n$--doped with
a multitude of different elements. This makes silicon the material of
choice for applications in electronics.

Silicon is also the leading player in the important field of
photovoltaic energy production. In fact, the large majority of the
first-generation solar cells, based on a single $p$-$n$ junction, use
bulk silicon as absorber layer. Its band-gap ($E_\text{g} = 1.12$\,eV
at room temperature~\cite{bludau1974}) lies in the optimal interval of
values to maximize energy conversion efficiency, according to the
Schockley-Queisser limit~\cite{shockley-queisser}. However, the fact
that the gap is indirect, and that the optical gap is larger than
3\,eV, makes silicon a very bad absorber of
sunlight~\cite{lautenschlager1987}. Therefore, the absorbing layer has
to be at least 100\,$\mu$m thick and, as a consequence, the crystal
needs to be very pure, so that the mean-free path of the carriers is
comparable with the size of the
layer~\cite{little1997,chopra2004}. This is a limiting factor in the
reduction of the cost of photovoltaic modules.

Due to these well-known limitations, materials with better absorption
coefficients in the visible have been put forward in the past years in
order to replace silicon in thin-film solar
cells~\cite{repins2008,rau1999,todorov2010,redinger2011}. Emphasis has
obviously been given to direct band-gap materials, with gaps between 1
and 1.5\,eV and with absorption spectra that strongly overlap with the
solar spectrum~\cite{vidal2010,botti2011,aguilera2011}. However, and
in spite of many spectacular advances, none of these materials has
been able to dethrone silicon. Of course, the ideal solution would be
to engineer silicon such that it absorbs strongly in the visible, in
order to use thin-film technology and at the same time maintain the
silicon-based processes that are currently employed. This would allow
for thinner, flexible, and cheaper silicon solar cells.

Aiming at maximizing the spectral overlap, some authors have recently
suggested different ways to manipulate the optical properties of
silicon. Nanostructuring is a largely explored way to obtain
direct-gap silicon by relaxing translational
symmetry~\cite{li2008}. However, the existence of a direct gap does
not guarantee that transitions at the absorption edge are dipole
allowed. Indeed, this is usually not the case in silicon
nanostructures~\cite{li2010}. D'Avezac {\it et
  al.}~\cite{d-avezac2012} proposed recently an ultrathin
silicon/germanium superlattice with excellent absorption properties.
The experimental synthesis of such system requires, however, control
of the growth of pure monolayers. Compensated doping of silicon obtained
through substitutional impurities was also shown to increase by 25\%
the absorption of 10\,$\mu$m thick silicon
layers~\cite{samsonidze2011}. Nevertheless, doping is expected to
introduce recombination centers detrimental for the performances of
the device.

Other works investigated the possibility to modify the band structure
of silicon by using allotropic phases with different crystal
symmetries~\cite{malone2008R,malone2008,malone2010,malone2012}. The
most stable phase of silicon, at ambient conditions, is the cubic
diamond structure. High pressure phases have been experimentally
studied up to 248\,GPa~\cite{duclos1990}, and several calculations of
the phase transitions are present in
literature~\cite{needs1995,pfrommer1997,mujica2003}. Upon increase of
pressure silicon exhibits a series of phase transitions: from cubic
diamond to $\beta$-Sn at around 12\,GPa, from $\beta$-Sn to
orthorhombic (Imma symmetry)~\cite{imma} and then to simple hexagonal
at 13--16\,GPa, from simple hexagonal to an orthorombic Cmca phase at
about 38\,GPa~\cite{hanfland1999}, from Cmca to hexagonal close pack
at 42\,GPa, and finally to face-centered cubic at 78\,GPa.  It has
been known for more than 20 years that, upon slow release of pressure
from the $\beta$-Sn structure, silicon transforms into the rombohedral
R8 and the body-centered BC8 crystals, characterized by distorted
sp$^3$ bondings\cite{needs1995,pfrommer1997,mujica2003}. R8 and BC8
are metastable phases, as they keep existing at ambient conditions. If
BC8 silicon is heated to temperatures in the range from 200 to
600\,$^{\circ}C$, it transforms to another metastable phase: the
lonsdaleite hexagonal diamond~\cite{lonsdaleite}.  Two other phases,
the so-called Si-VIII and Si-IX structures, were observed
experimentally upon rapid release of pressure from $\beta$-Sn
silicon~\cite{zhao1986}.  However, their crystal lattices were not
fully characterized.  Nanoindetation experiments gave evidences of
another phase, known as Si-XIII~\cite{Si-XIII}, although also in this
case very little information is available on its crystal
structure. Other meta-stable phases that lay close in energy to cubic
diamond silicon have been proposed theoretically, using a multitude of
methods~\cite{BCT,ST12,wang2010,malone2012}.  Malone {\it et al.}
calculated the band structure and the absorption spectrum one of these
low-energy arrangements, charachterized by a body-centered-tetragonal
(BCT) unit cell~\cite{malone2010}. A similar study was carried out by
the same authors also for the experimental R8
structure~\cite{malone2008R}.  Unfortunately, BCT silicon did not show
enhanced absorption properties in the visible if compared to cubic
diamond silicon and R8 turned out to have a too small gap (0.24\,eV)
for photovoltaic applications.  However, the low-pressure phase
diagram of silicon is still relatively unexplored, which makes us
believe that there might exist unexplored silicon phases with optical
properties suitable for thin-film photovoltaics.

In this paper we present the results of a structural prediction search
that leads to a number of previously unknown, low-energy, $sp^3$
phases of silicon that have excellent properties for photovoltaic
applications, with quasi-direct band-gaps between 1--1.5\,eV, and
excellent absorption properties for solar light. Note that the band
gap imposes an upper bound on the open-circuit voltage of the device,
implying that too small indirect band-gap compounds should be
discarded. All structures we selected have a total energy
higher than the one of cubic silicon by less than 0.15\,eV per atom.
This restriction assures that the experimental synthesis of these structures 
is energetically plausible.

The silicon crystalline arrangements under investigation were found by
performing a structural relaxation of low-enthalpy carbon primitive
cells with 8 atoms, which we had previously
calculated~\cite{prl_Z-carbon_2012} by applying a very efficient
structural prediction algorithm, the minima hopping method
(MHM)~\cite{goedecker_2004,amsler_2010}.  The MHM was designed to
obtain the low-enthalpy structures of a system given solely its
chemical composition. The enthalpy surface is explored by performing
consecutive short molecular dynamics escape steps followed by local
geometry relaxations taking into account both atomic and cell
variables. The initial velocities for the molecular dynamics
trajectories are chosen approximately along soft mode directions, thus
allowing efficient escapes from local minima and aiming towards low
energy structures. The predictive power of this approach has already
been demonstrated for a wide range of
applications~\cite{hellmann_2007,roy_2009,bao_2009,willand_2010,prl_Z-carbon_2012,prl_disilane_2012}.
As carbon admits both $sp^3$ and $sp^2$ bonds, its phase diagram is
much richer than the one of silicon. For this reason the low-enthalpy
carbon allotropes found at low pressure provide a good structural
database for silicon analogues. Indeed, we observed that the carbon
structures with $sp^2$ bonds turn out to be unstable and relax into a
different geometry with $sp^3$ arrangement.

\begin{figure}[t]
  \subfigure[ M-10]{\includegraphics[width=0.3\columnwidth,angle=0]{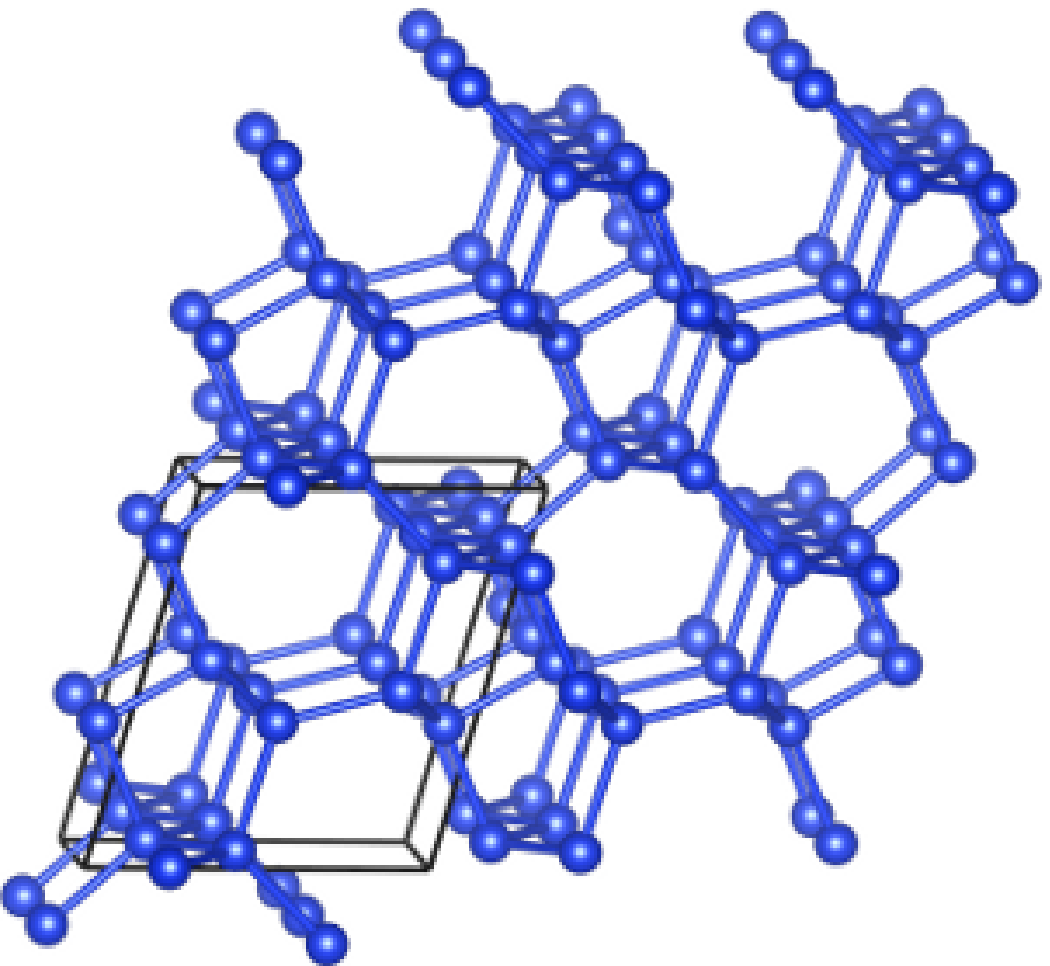}} 
  \subfigure[ C222$_1$]{\includegraphics[width=0.3\columnwidth,angle=0]{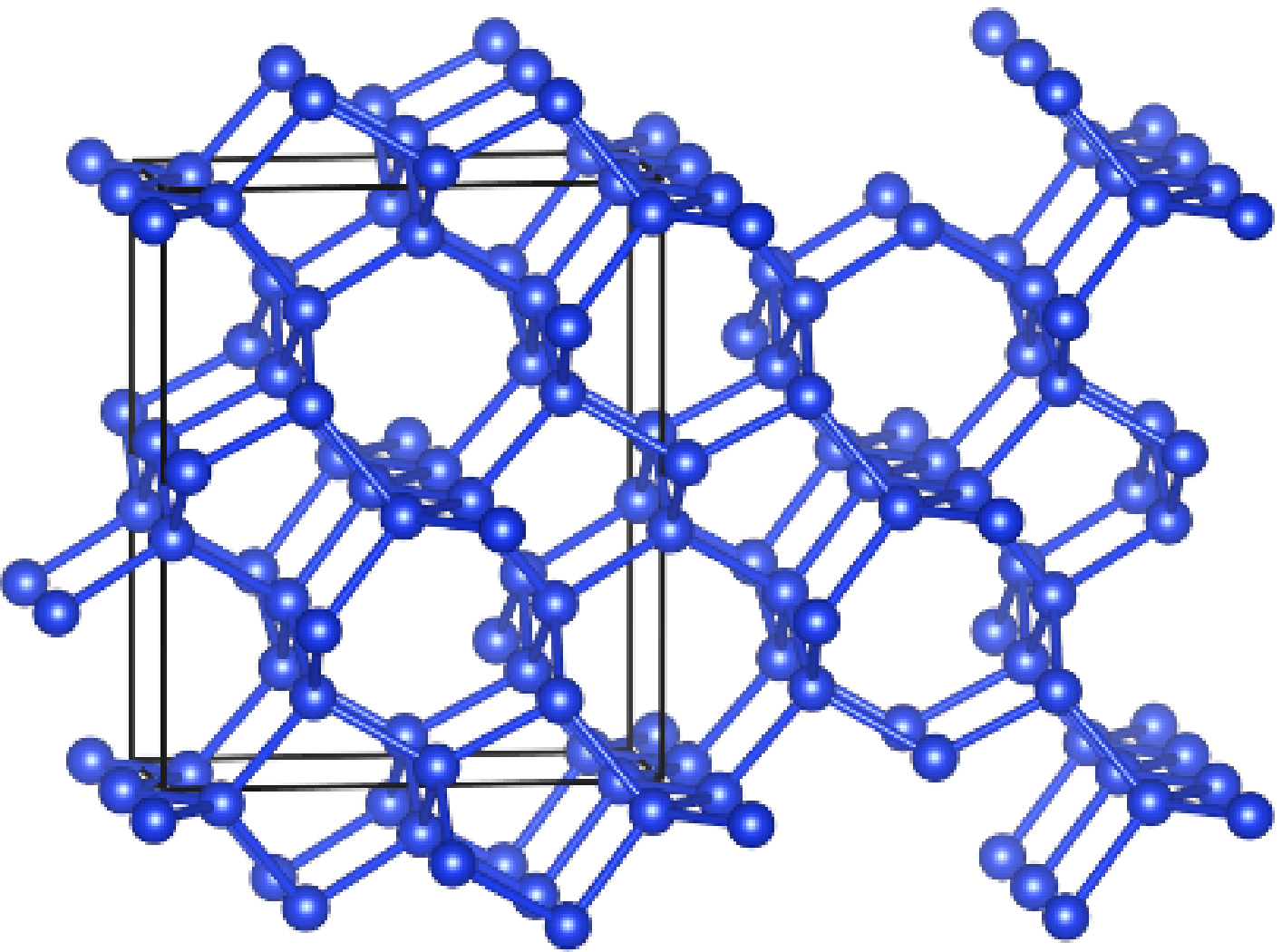}}
  \subfigure[ Imma$^{(2)}$]{\includegraphics[width=0.3\columnwidth,angle=0]{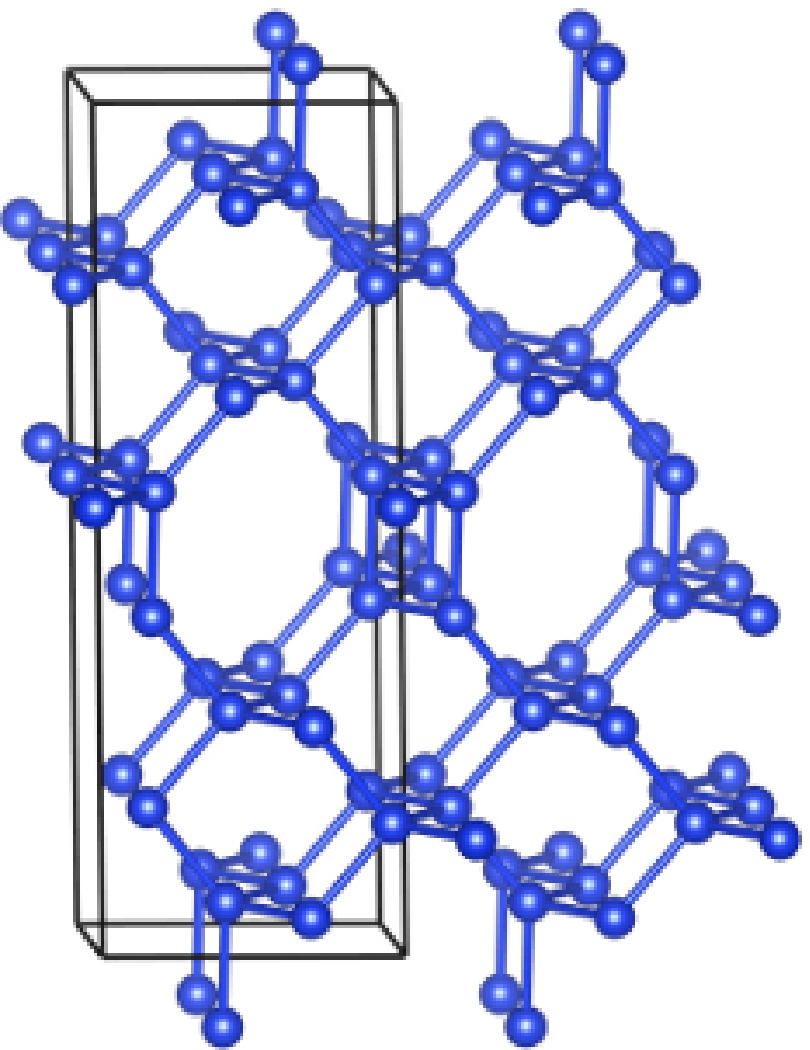}}
  
  \subfigure[ Cmcm]{\includegraphics[width=0.3\columnwidth,angle=0]{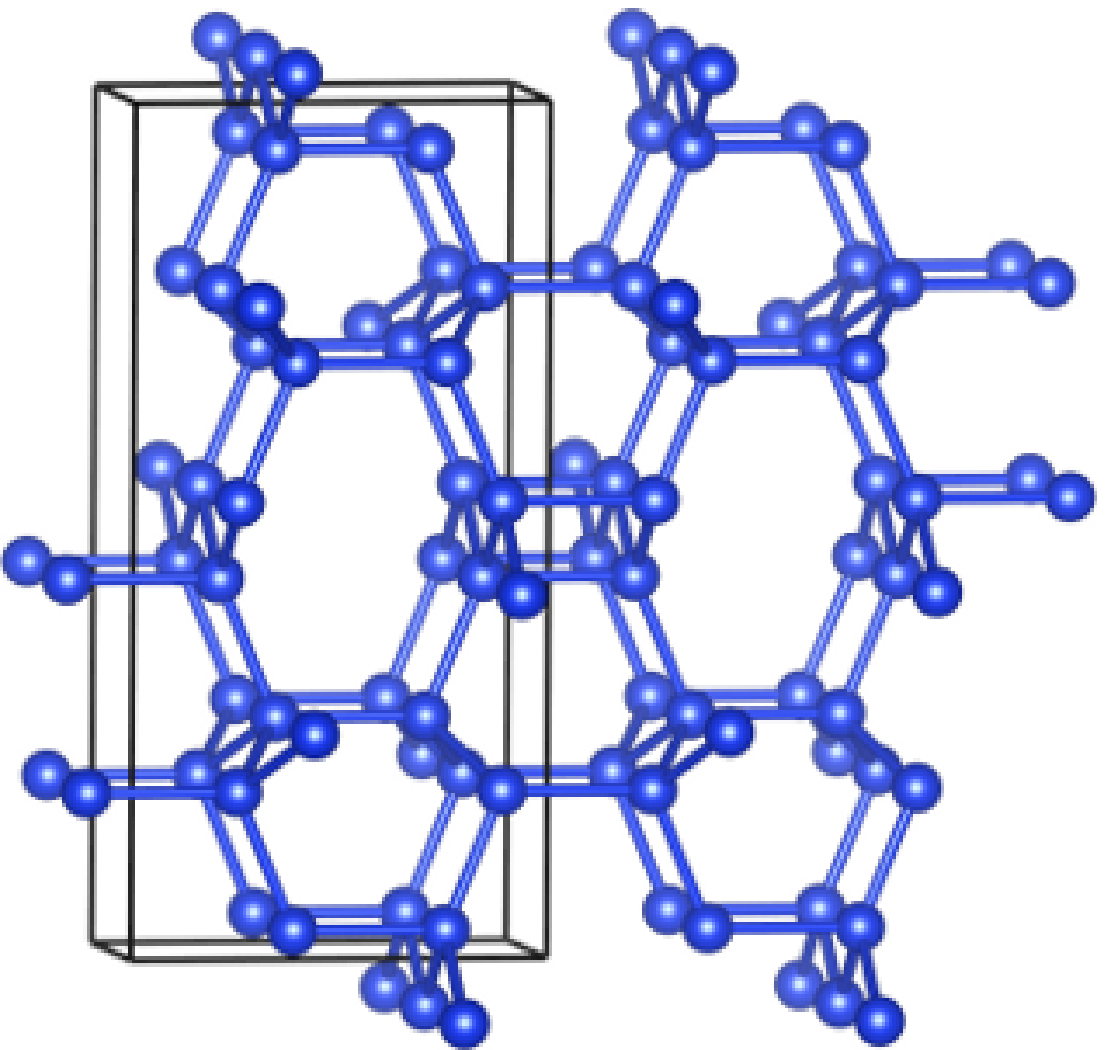}}
  \subfigure[ P-1]{\includegraphics[width=0.3\columnwidth,angle=0]{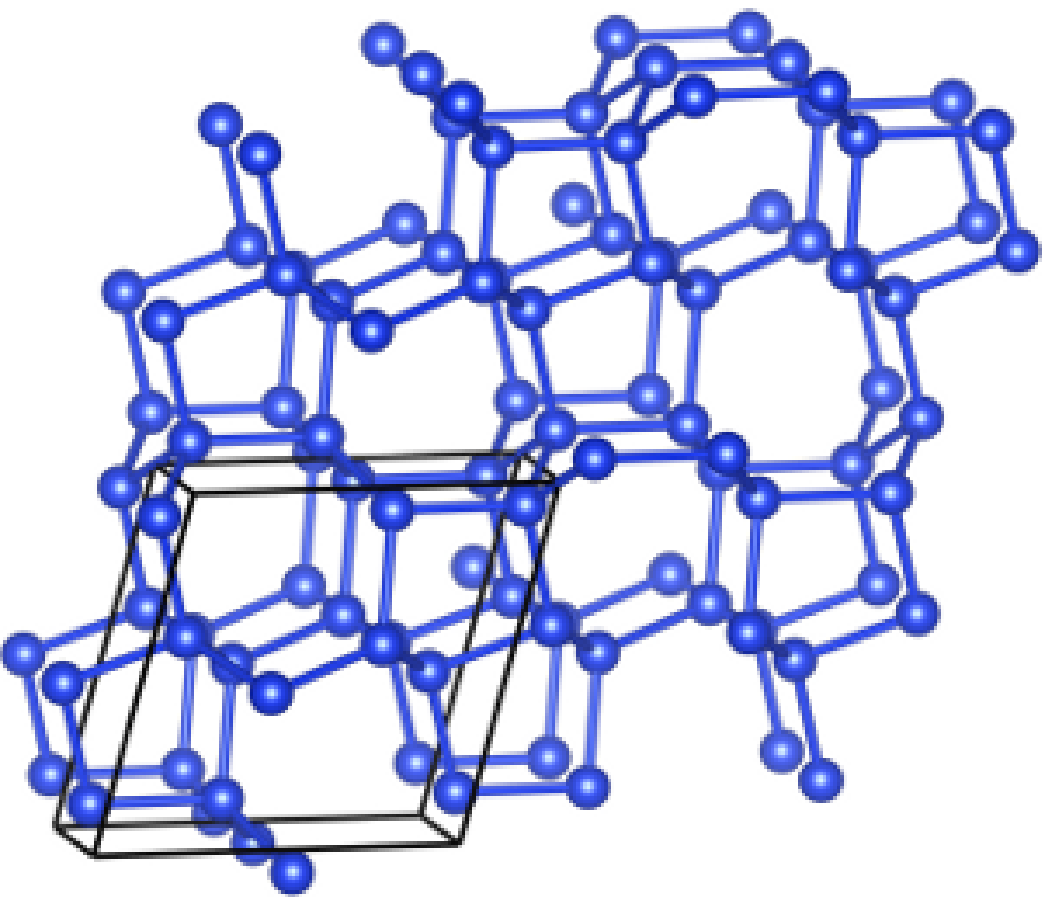}}
  \subfigure[ P2$_1$/c]{\includegraphics[width=0.3\columnwidth,angle=0]{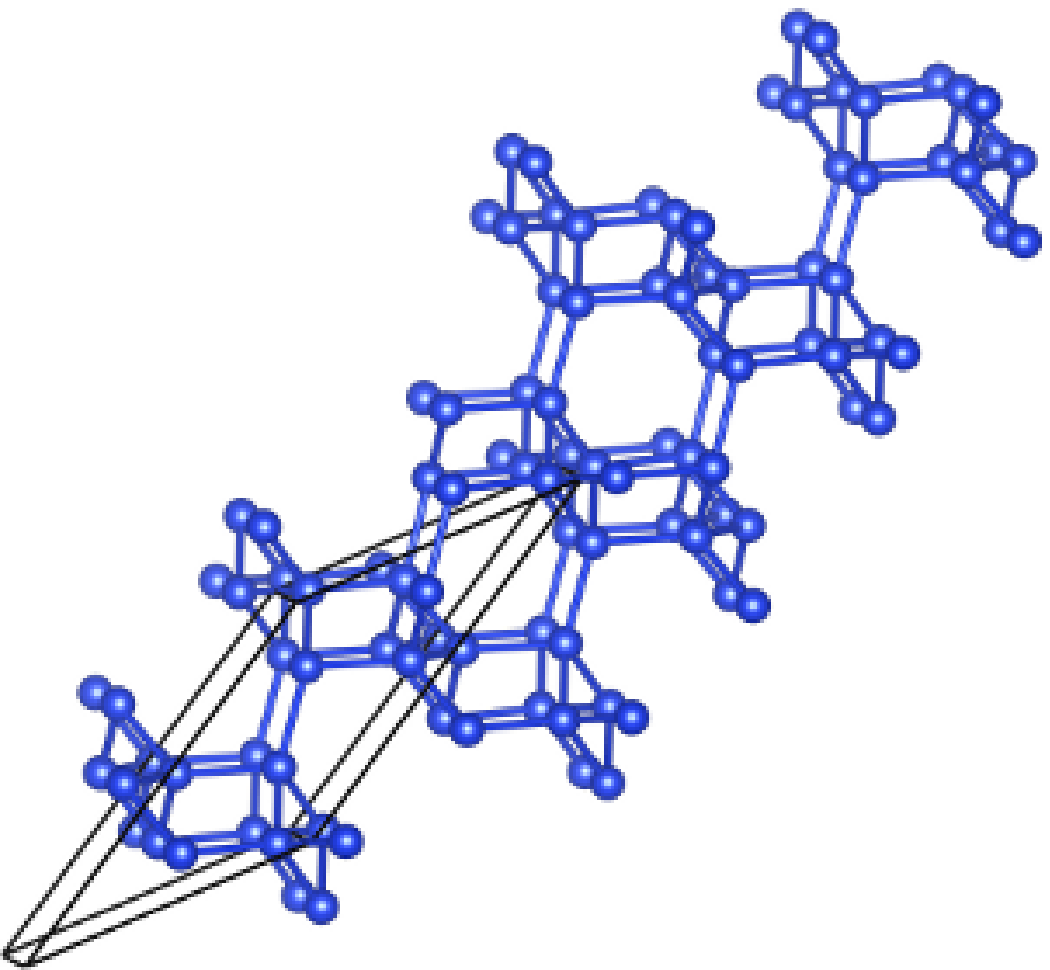}}
  \caption{Representation of the most important $sp^3$ Si structures
    found in this work. The superscript in Imma$^{(2)}$ serves to
    distinguish this structure from the high-pressure phase Si-XI of
    the same symmetry.}
  \label{fig:structures}
\end{figure}

\begin{table}[t]
  \caption{Selected low-energy silicon allotropes found in this study. 
    Total energies per atom are given relatively to the diamond
    structure.  PBE indirect gaps are compared with corresponding $GW$
    indirect gaps. All energies are in eV. The references to works that
    present carbon analogues are given in the last column.
    \label{tab:structures}}

  \begin{tabular}{rccccl}
    structure & spc group & $E_\text{total}$/atom & $E_\text{gap}^\text{PBE}$ & $E_\text{gap}^{GW}$ & C analogue \\
    \hline
    Z        & 65 & 0.06 & 0.72 & 1.22 & Ref.~\protect\onlinecite{prl_Z-carbon_2012,PRBR_C222,PRL_Zcarbontwo} \\    
    M        & 12 & 0.07 & 0.46 & 0.99 & Ref.~\protect\onlinecite{M12carbon_Oganov} \\
    M-10     & 10 & 0.08 & 0.97 & 1.40 &  Ref.~\protect\onlinecite{Cmcm_PRB,prb_M10_carbon_2012} \\
    C222$_1$ & 20 & 0.08 & 0.90 & 1.39 & Ref.~\protect\onlinecite{PRBR_C222}   \\
    Imma$^{(2)}$ & 74 & 0.12 & 0.63 & 1.15 & --  \\
    Cmcm     & 63 & 0.13 & 0.32 & 0.83 & Ref.~\protect\onlinecite{Cmcm_PRB} \\
    P-1      &  2 & 0.13 & 0.28 & 0.75 & --  \\
    P2$_1$/c & 14 & 0.14 & 0.73 & 1.29 & --
  \end{tabular}
\end{table}

The evaluation of energies and forces required for the MHM were
performed within density functional theory (DFT) using the {\sc ABINIT}
code~\cite{abinit}, with simulation cells of 8 carbon atoms at 15~GPa.
The lowest energy silicon analogues were further relaxed and 
characterized using the {\sc VASP} 
code~\cite{vasp}. We selected our
$k$-point grids to ensure an accuracy of 0.01\,eV in the total energy,
and all forces were converged to better than 0.005\,eV/\AA. To approximate
the exchange-correlation functional of DFT we used the
Perdew-Burke-Ernzerhof (PBE)~\cite{PBE} generalized gradient
approximation.

In an energy interval of 0.15\,eV per atom (around 1740\,K) from the
diamond structure, and ignoring the well-known cubic diamond and
hexagonal diamond silicon, we found a total of 16 structures. We
checked the stability of these phases by calculating their phonon band
structure, in order to make sure that all phonon modes are real. We
note that the experimentally known meta-stable phases of silicon, R8
and BC8, have higher energies than the phases considered
here. Interestingly, also the most studied theoretical meta-stable
phases (ST12, BCT, Ibam etc.) are energetically less favored than the
new structures presented in this work.

All our structures have $sp^3$ bonding, and are semiconducting with
Kohn-Sham PBE (indirect) gaps ranging from 0.3 to 1.2\,eV. Some of
the carbon analogues can
already be found in the literature (see Table~\ref{tab:structures}). It
is noticeable that none of the around 300 semiconducting structures
identified by our simulation~\cite{footnote} has a direct
quasi-particle gap. Nevertheless, several of them exhibit a
quasi-direct gap.  The lowest-energy structures we found with
promising quasiparticle gaps are listed, together with their space
group number and the value of the calculated gaps, in
Table~\ref{tab:structures}. The corresponding atomic arrangements are
depicted in Fig.~\ref{fig:structures}.  

\begin{figure}[t]
  \includegraphics[width=0.99\columnwidth,angle=0]{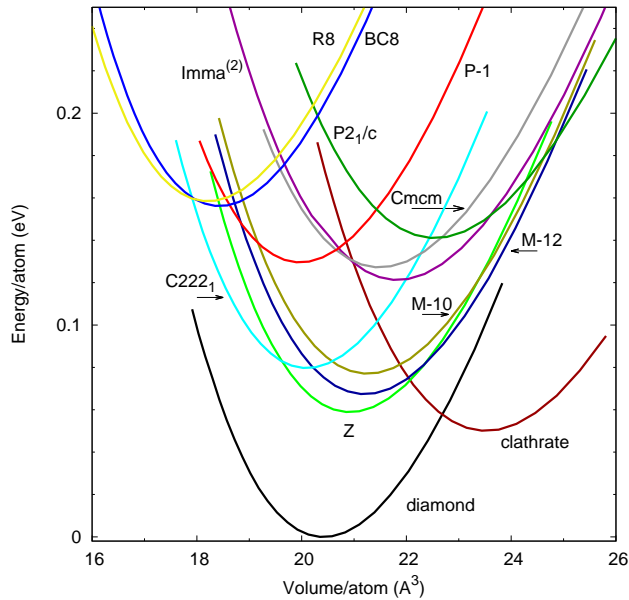}
  \caption{(Color online.) Energy per atom as a function of volume per
    atom for the most important sp$^3$ silicon structures found in
    this work. The zero of energy is the total energy per atom of the
    cubic diamond structure in equilibrium. For comparison we also
    include some already known structures of silicon, e.g., cubic
    diamond, a clathrate~\cite{clathrate}, BC8, R8.}
  \label{fig:energy}
\end{figure}

In Fig.~\ref{fig:energy} we show the total energy per atom as a function of 
volume per atom for the new silicon allotropes found in this study. 
For the sake of comparison we included the energy curves of some well known phases of silicon, e.g., 
cubic diamond silicon, a clathrate, the BC8 and R8 structures.

By inspection of Fig.~\ref{fig:energy} and Table~\ref{tab:structures} we can easily deduce 
that, among the novel silicon structures, the lowest energy allotropes is
Z-silicon, followed by M and M-10 silicon. These crystal structures
have carbon counterparts that are thought to play an important role in
cold-compressed graphite~\cite{prl_Z-carbon_2012}. The volumes per atom at equilibrium of the
new structures are between 20 and 22\,\AA$^3$, which places them in an
intermediate position between the open clathrates and the
experimentally observed BC8 and R8 structures. 

\begin{figure}[t]
  \includegraphics[width=0.99\columnwidth,angle=0]{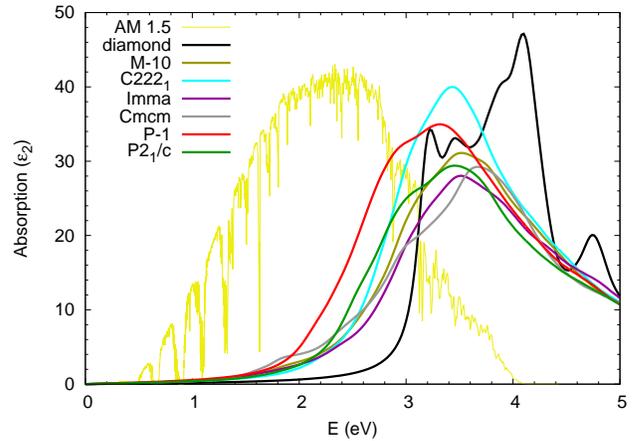}
  \caption{(Color online.) Absorption spectra of the most important sp$^3$ Si
    structures found in this work compared to the reference air mass
    1.5 solar spectral irradiance~\cite{AM1.5sun}, given in arbitrary units.}
  \label{fig:spectra}
\end{figure}

In order to characterize the fundamental and optical band gaps,
together with the absorption spectra, we performed perturbative
$GW$~\cite{hedin} calculations and we solved the Bethe-Salpeter
equation~\cite{BSE,lucia1998} using the {\sc abinit} package~\cite{abinit}. We
want to stress that we employed the most accurate methods available in
the community to study electronic excitations, that have proved to
give excellent results for a wide range of $sp$ semiconductors and
insulators.~\cite{aulbur,review-lucia} In particular, the BSE
absorption spectrum of cubic diamond silicon ~\cite{lucia1998} is in
excellent agreement with experimental measurements~\cite{lautenschlager1987},
thanks to the inclusion of excitonic effects.  We used
Troullier-Martins norm-conserving pseudopotentials~\cite{TM} and the
PBE exchange-correlation functional~\cite{PBE}. The planewave cut-off
energy for all runs was 15\,Hartree. For $GW$ calculations the cutoff
of the dielectric matrix was set to 5\,Hatree and we needed 12 states
per atom, together with the method of Bruneval and Gonze, to achieve a
convergence of around 0.05\,eV in the band gap. For the Bethe-Salpeter
calculations we used a cutoff of 2.5\,Hartree for the dielectric
matrix and a k-point spacing of around 0.012 in reciprocal lattice
vector units for the shifted k-point mesh. This criterion yields a
14$\times$14$\times$14 shifted grid for the diamond structure.

All the $GW$ corrections to the Kohn-Sham PBE band structures can be
approximated by a rigid shift of the conduction bands, as they are fairly independent of the k-points. 
Furthermore, this rigid shift does not change
significantly among the different crystal lattices, ranging from 0.4
to 0.6\,eV, even if the Kohn-Sham gaps exhibit larger variations.  By
inspecting the values for the indirect gaps in
Table~\ref{tab:structures} we can observe that several silicon
structures have a gap in the optimal frequency interval for maximizing
photovoltaic efficiency. However, one should not forget that all the
gaps that we calculated are indirect, even if in many cases the direct
gaps are much smaller than in the cubic diamond phase. This fact
suggests that it is conceivable that the absorption edge for direct
transitions of these metastable allotropes is at significantly lower
energies than in conventional silicon.  However, information on the
band structure alone is not conclusive, since the dipole matrix
elements between states close to the valence band maximum and the
conduction band minimum can be very small and suppress light
absorption close to the absorption edge. In view of that, we performed
accurate calculations of the absorption spectra by solving the
Bethe-Salpeter equation.

In Fig.~\ref{fig:spectra} we show the calculated absorption spectra of
the allotropes that are most promising for applications in solar
cells. The calculated absorption spectrum of the cubic diamond phase
is also shown for comparison, together with the reference air mass
(AM) 1.5 solar spectral irradiance. Note that indirect absorption
contributions, which arises from phonon assisted interband transitions
are not included in our calculations. All new structures start to
absorb very close to their direct gap, with the result that their
absorption spectra overlap significantly with the solar spectrum.
Remarkably, the absolute optical absorption between 1.5 and 3\,eV
strictly is comparable to that of Cu(In,Ga)Se$_2$ compounds, that are
regarded as excellent absorbers for thin-film solar cell
technology~\cite{exp-abs-CIGS}. This indicates that these new silicon phases 
could potentially be employed in highly efficient silicon thin-film solar panels.

If we make a critical evaluation of all calculated properties, we can
conclude that the most promising allotropes are the M-10 and the
C222$_1$. In fact, their total energies per atom are particularly low,
their indirect gap is 1.4\,eV, while the direct gap is only slightly
larger (1.5\,eV for M-10 and 2.0\,eV for C222$_1$).  Moreover,
absorption is dipole allowed starting from the absorption edge, as it
is revealed by the long tails of their BSE absorption spectra in the
visible.

In conclusion, we predict several novel metastable phases of silicon,
obtained through a structural search based on carbon analogues found
using the minima hopping method. We filtered the most promising
structures for use in thin-film photovoltaics by imposing that the
total energy per atom does not differ by more than 0.15\,eV from the
total energy per atom of cubic diamond silicon. Moreover, we imposed
that the fundamental band gap is in the range 1.0--1.5\,eV. Some of
the new structures revealed a strong absorption in the visible, with
absorption coefficients comparable to those of chalcopyrite absorbers
used in thin-film record solar cells. These results call for further
experimental studies of the low-pressure phase diagram of silicon, and
could open new routes to design highly efficient thin-film silicon
solar cells.

JAFL acknowledges the CONACyT-Mexico. MALM acknowledges support from
the French ANR (ANR-08-CEXC8-008-01). Financial support provided by
the Swiss National Science Foundation is gratefully acknowledged.
Computational resources were provided by IDRIS-GENCI (project
x2011096017) in France and the Swiss National Supercomputing Center
(CSCS).

\end{document}